\documentclass[prl,aps,twocolumn]{revtex4}  
\usepackage{amsmath,epsfig}  
  
\begin{document}  
% ---------------------------------------------------------------------%  
%  
%  
%-------------------------------------------------------------  
%  
%  
\title{Lattice Models with ${\cal N}=2$ Supersymmetry}
  
\author{
Paul Fendley$^1$, Kareljan Schoutens$^2$ and Jan de Boer$^2$ 
\medskip}  
\affiliation{  
$^1$ Department of Physics, University of Virginia,   
Charlottesville, VA 22904-4714 USA \\  
{\tt fendley@virginia.edu}\smallskip \\   
$^2$ Institute for Theoretical Physics, University of Amsterdam,
Valckenierstraat 65, 1018 XE Amsterdam, The Netherlands \\  
{\tt kjs@science.uva.nl, jdeboer@science.uva.nl}}
  
\begin{abstract}  
\smallskip  
We introduce lattice models with explicit ${\cal N}$=2 supersymmetry.
In these interacting models, the
supersymmetry generators $Q^\pm$ yield the Hamiltonian $H=
\{Q^+,Q^-\}$ on any graph. The degrees of freedom can be described as
either fermions with hard cores, or as quantum dimers. The Hamiltonian
of our simplest model contains a hopping term and a repulsive
potential, as well as the hard-core repulsion. We discuss these models
from a variety of perspectives: using a fundamental relation with
conformal field theory, via the Bethe ansatz, and using cohomology 
methods. The simplest model provides a 
manifestly-supersymmetric lattice regulator for the supersymmetric point
of the massless 1+1-dimensional Thirring (Luttinger) model. We
discuss the ground-state structure of this same model on more
complicated graphs, including a 2-leg ladder, and discuss some
generalizations.
  
\end{abstract}  
  
  
\date{October 16, 2002}  
  
\maketitle  
  
Supersymmetry is an exceptionally powerful theoretical tool. It often
allows exact computations in field theory and string theory, even when
interactions are strong. In this paper we introduce ${\cal N}=2$
supersymmetric lattice models describing interacting fermions and
monomer-dimer systems. We show that the continuum limit
of the simplest of these models, defined on a one-dimensional lattice,
is a well-known $1+1$-dimensional quantum field theory with ${\cal N}=2$
superconformal symmetry.
  
Our definition of ${\cal N}$=2 supersymmetry is that
the Hamiltonian $H$ is built from
two nilpotent fermionic generators denoted 
$Q^+$ and $Q^-=(Q^+)^\dagger$ \cite{Witten}. It is
\begin{equation}H=\{Q^+,Q^-\}.
\label{QQH}  
\end{equation}
The fact that $Q^+$ and $Q^-$ commute with $H$
follows from the nilpotency $(Q^+)^2=(Q^-)^2 =0$.
Our models also have a
fermion-number symmetry generated by $F$ with $[F,Q^\pm]=\pm
Q^\pm$.  We shall show how, in at least some cases, this lattice
supersymmetry extends to a space-time super(conformal) symmetry in the
field theory describing the continuum limit. Lattice
models with a symmetry involving fermionic generators, such as the
$t$-$J$ model at $J$=$\pm 2t$, are often called ``supersymmetric'' in
the condensed-matter literature, but do not have a Hamiltonian of the
form eq.~(\ref{QQH}).

All eigenvalues $E$ of the Hamiltonian eq.~(\ref{QQH}) satisfy
$E\ge 0$. All eigenstates form either singlet or doublet representations  
of the supersymmetry algebra.  All states $|g\rangle$ with $E=0$ must  
be singlets: $Q^+|g\rangle=Q^-|g\rangle=0$ \cite{Witten}.  Conversely,  
all singlets must have $E=0$.  All the other eigenstates of $H$ can be  
decomposed into doublets under the supersymmetry, and conversely any  
doublet representation is an eigenstate. This is simple to prove: a  
doublet consists of two states $|s\rangle,Q^+|s\rangle$, where  
$Q^-|s\rangle =0$. It follows from the definition of $H$ and the  
nilpotency of $Q^\pm$ that both of these states are eigenstates of $H$  
with the same eigenvalue. All eigenstates can be decomposed into  
doublets:  the four-dimensional  
representation $(|s'\rangle,Q^-|s'\rangle,Q^+|s'\rangle, Q^+Q^-|s'\rangle)$
is reducible. Let
$$|s\rangle\equiv |s'\rangle -\frac{1}{E_s} Q^+Q^-|s'\rangle$$
where $E_s>0$ is defined by $H|s'\rangle = E_s|s'\rangle$. Then
$Q^-|s\rangle =0$, and $(|s\rangle,Q^+|s\rangle)$ and
$(Q^-|s'\rangle,Q^+Q^-|s'\rangle)$ form two irreducible doublets.
  
The models we introduce can be defined on any lattice
(or actually, any graph) in any dimension. The simplest model involves
a single species of fermion $c_i$, placed at any site $i$ of the
lattice. The fermion obeys the usual anticommutator
$\{c_i,c^\dagger_j\}=\delta_{ij}$, and the operator $F=\sum_i
c^\dagger_i c_i$ counts the number of fermions.  We impose the
restriction that the fermions have hard cores, meaning that fermions
are not allowed on neighboring sites.
A hard-core fermion is created by
$c^\dagger_i {\cal P}_{<i>}$, where the projection
operator ${\cal P}_{<i>}$ requires all sites
neighboring $i$ to be empty:
\begin{equation}  
{\cal P}_{<i>}= \prod_{j\hbox{ next to } i} (1-c^\dagger_j c_j) \ .  
\end{equation}  
On this space of states, the supersymmetry
operators are defined by
\begin{equation}  
Q^+=\sum_i c^\dagger_i {\cal P}_{<i>} \qquad\quad  
Q^-=\sum_i c_i {\cal P}_{<i>}.  
\label{susydef}  
\end{equation}  
It is easy to verify that $(Q^+)^2=(Q^-)^2=0$.
The Hamiltonian is therefore
\begin{equation}  
H= \sum_i \sum_{j\hbox{ next to }i} {\cal P}_{<i>} c^\dagger_i   
c_j {\cal P}_{<j>} + \sum_i {\cal P}_{<i>}.
\label{ham}  
\end{equation}
The first term in the Hamiltonian allows fermions to hop to
neighboring sites, with the projectors maintaining the hard-core
repulsion. The second term favors having more fermions, as long
as they are more than two sites from each other. Thus one can
view it as a repulsive potential for fermions, 
in addition to the hard core.
  
There are two key questions to try to answer. The first is:
what properties can be computed exactly using the supersymmetry? 
We have already noted the positive energy and 
the pairing in the excited-state spectrum,
but these are just the simplest consequences of the supersymmetry.
The second question is: what (if any) field theory
describes the model in the continuum limit?

To illustrate the power of supersymmetry, we find the ground states
for a chain of six sites and periodic boundary conditions.  First, we
count all the states. There is one state $|0\rangle$ with $f$=0, and
six states $c_i^\dagger|0\rangle$ with $f$=1, while because of the
hard cores, there are nine states with $f$=2, and two with $f$=3. The
vacuum obeys $Q^-|0\rangle= 0$ and $Q^+|0\rangle =\sum_{i=1}^6
c_i^\dagger|0\rangle$, so $(|0\rangle,Q^+|0\rangle)$ make up a
doublet. The remaining five states with $f$=1 are all annihilated by
$Q^-$, and $Q^+$ acts non-trivially on them. There are thus
five doublets with $(f,f+1)=(1,2)$. The states with $f$=3 are both
annihilated by $Q^+$, and $Q^-$ acts non-trivially on both, giving two
doublets with $(f,f+1)=(2,3)$. This accounts for all the states in the
theory, except for two states with $f$=2.  These two cannot form a
doublet, because they have same fermion number.  They therefore must
be singlets, so there are two $E$=0 ground states in this theory, both
with $f$=2.  With a little more work, one
finds that they have eigenvalues $\exp(\pm i\pi/ 3)$ under translation
by one site. 

A basic quantity in a supersymmetric theory is the Witten index~\cite{Witten}  
\begin{equation}
W = \hbox{tr}\left[(-1)^F e^{-\beta H}\right]. \label{Windex}
\end{equation}
Because the two states in a doublet have the same energy, their
contribution to $W$ cancels, leaving the trace only over ground states
and $W$ independent of $\beta$.  $W$ can thus be found by evaluating
(\ref{Windex}) in the $\beta \to 0$ limit, where all states contribute
with weight $(-1)^F$.  For example, for the six-site chain discussed
above, we confirm that $W=1-6+9-2=2$. For our model on the cube, one
finds $W=1-8+16-8+2=3$.

Computing $W$ for the model eq.~(\ref{ham}) on a general graph poses a
fascinating combinatorial problem.  Cohomology theory is a powerful
tool to compute the number of ground states at any fermion number, and
therefore also $W$. The supersymmetry generator $Q^+$ satisfies
$(Q^+)^2=0$, and the $E=0$ ground states are precisely the states
$|s\rangle$ that satisfy $Q^+|s\rangle = 0$ and that cannot be written
in the form $|s\rangle=Q^+|s^\prime\rangle$. Those states form what is
called the cohomology of the operator $Q^+$ and for its computation a
variety of techniques are available. These include the `spectral
sequence' technique, which can be applied in this context as follows:
one splits the lattice in two sublattices, with corresponding fermion
number operators $F_1$ and $F_2$, so that $F=F_1+F_2$. $Q^+$ can also
be split as $Q^+_1+ Q^+_2$, so that $Q_i^+$ increases $F_i$ by
one. The two operators $Q^+_1$ and $Q^+_2$ are nilpotent and
anti-commute. The first step in the spectral sequence is to compute
the cohomology of $Q^+_1$. $Q^+_2$ becomes an operator acting on this
cohomology, and the second step is to compute the cohomology of
$Q^+_2$ on this subspace.  Often the process terminates here, and the
result is the cohomology of $Q^+$. In general the procedure continues
for a finite number of additional steps, see e.g.~\cite{botttu} for
details.  A similar procedure exists for a decomposition of a lattice
into $n$ sublattices, and in particular when every sublattice consists
of a single site.  Applying this to the $N$-site periodic chain, with
$F_1$ consisting of every third site, we find ground states solely at
fermion number $f=\hbox{int}((N+1)/3)$. For a chain with $N=3p$ with
$p$ integer, we find two ground states (so $W=2(-1)^p$), while for
$N=3p\pm 1$, there is a single ground state (so $W=(-1)^p$).

We now address our questions in the one-dimensional case,
where the Hamiltonian eq.~(\ref{ham}) is on an $N$-site chain: 
\begin{equation}  
H= \sum_{i=1}^N \left[P_{i-1}  
\left( c^\dagger_i c_{i+1} + c^\dagger_{i+1}c_i\right)P_{i+2}  
 +P_{i-1}P_{i+1}\right]  
\label{ham1}  
\end{equation}  
where $P_i\equiv 1-c^\dagger_i c_i$ is the projector on a single site.  
We take periodic boundary conditions, so indices are defined 
mod $N$. The translation operator $T$ commutes with
both $H$ and $F$, and its
eigenvalues $t$ satisfying $t^N=1$ 
characterize the eigenstates of $H$. The Hamiltonian eq.~(\ref{ham1})
resembles a lattice version of the Thirring model,  
a 1$+$1-dimensional field theory with a four-fermion interaction term.   
Below we will make the connection precise.
  
Before we apply the Bethe ansatz,  
we indicate how the space of states ${\cal H}_N$ of the model can be  
obtained by applying a systematic `finitization' procedure  
\cite{kjs-bs} to the chiral spectrum of a specific ${\cal N}=2$   
superconformal field theory (SCFT) with central charge $c=1$.   
This construction follows two steps. In the first step,  
the full chiral Hilbert space of the SCFT is written in a   
`quasi-particle' basis, with the fundamental quasi-particles 
forming a supersymmetry doublet with charges $(1/3,-2/3)$. 
In the second step, the momenta of the individual quasi-particles 
are constrained to  
a maximum value in the order of $N$, corresponding to a discretization  
of the space direction of the SCFT with spacing of the order $1/N$.   
This then leads to a truncated or `finitized' partition sum $Q_N(q,w)$,  
which is closely related to the partition sum   
$Z_N(q,w) = {\rm Tr}(q^{\frac{N}{2\pi i} \log T} \, w^{(N-3F)})$,   
which keeps track of the eigenvalues of the translation operator $T$   
and the fermion number $F$ of the lattice model eq.~(\ref{ham1}).  
This entire construction, to be detailed elsewhere, respects  
the supersymmetry. This gives a rationale for the existence   
of supersymmetry generators on the space ${\cal H}_N$, and it  
provides an independent way of determining the quantum numbers   
$t$ and $f$ of the supersymmetric ground states. We   
remark that an analogous construction based on the spinon basis of   
the simplest $SU(2)$-invariant CFT naturally leads to the space of   
states of the Heisenberg and Haldane-Shastry models for $N$ spin-1/2   
degrees of freedom. Clearly, many generalizations are possible.  
  
Having discussed the model eq.~(\ref{ham1}) from different points of
view, we now give
some of the results of a Bethe ansatz computation \cite{FSii}.
This computation lets us identify the continuum limit of the
theory. An eigenstate of $H$ with $f$ fermions is of the form  
\begin{equation}  
\phi^{(f)} = \sum_{\{i_k\}} \varphi(i_1,i_2,\dots i_f)  
c^\dagger_{i_1}c^\dagger_{i_2}\dots c^\dagger_{i_f}|0\rangle   
\label{Phidef}  
\end{equation}  
where we order $1\le i_1<i_2-1 <i_3-2 \dots$.   
Bethe's ansatz for the eigenstates of $H$ is \cite{Bethe}
\begin{equation}\varphi(i_1,i_2,\dots i_f) = \sum_P  
A_P\  \mu_{P1}^{i_1-1}\mu_{P2}^{i_2-1}\dots \mu_{Pf}^{i_f-1}.  
\label{phidef}  
\end{equation}  
for some numbers $\{\mu_1,\dots,\mu_f\}$ and $A_P$; the sum is over  
permutations $P$ of the set $(1,2,\dots,f)$. By construction, the
translation operator $T$ has eigenvalue $t=\prod_{k=1}^{f}\mu_k^{-1}$.

The next step is to find highest-weight states under the symmetries of 
the model. In  
Bethe's case, the symmetry is the $O(3)$ of the Heisenberg spin  
chain; here it is the supersymmetry. Here, $Q^-\phi^{(f)}=0$ 
requires
\begin{equation}  
t^{-1} (\mu_k)^{N-f}  = \prod_{j=1}^f   
\frac{\mu_k\mu_j +1 - \mu_k}{\mu_k\mu_j +1 -\mu_j}  
\label{bethe}  
\end{equation}  
for all $k=1\dots f$. 
These
are very similar to the Bethe equations for the antiferromagnetic
XXZ spin chain at $\Delta=\pm 1/2$ \cite{Baxter}.  The only
difference in (\ref{bethe}) is in the left-hand-side, which in the XXZ
case reads $(\mu_k)^N \tau$, where $\tau\ne 1$ corresponds to twisted
boundary conditions.

Demanding that a state be an eigenstate of a continuous symmetry
fixes all the free parameters in the Bethe
ansatz. The miracle of the ansatz is that in the Heisenberg 
and other integrable models, this state is also an eigenstate of the
Hamiltonian. In our case, supersymmetry provides the miracle. 
Any Bethe ansatz state obeying $Q^-|s\rangle=0$ is either a singlet
or part of a doublet, and so it must be an eigenstate of $H$. 
Its energy is
\begin{equation}  
E= N-2f + \sum_{k=1}^f \left[\mu_i + \frac{1}{\mu_i}\right].  
\label{energy}  
\end{equation}  
The supersymmetry doublets appear naturally within  
the Bethe ansatz. If ($\mu_1,\dots \mu_f$) satisfies the Bethe  
equations, then the set ($1,\mu_1,\dots\mu_f$) also must satisfy the  
Bethe equations, and moreover, both sets have the same energy $E$.  It  
is straightforward to check that the states associated with these two
sets are related by $\phi^{(f+1)} = Q^+ \phi^{(f)}$.
  
The Bethe equations are $f$ coupled polynomial equations of order $N$.  
They cannot be solved in closed form, and to make further progress,  
one usually needs to take $N$ large. In our case, however, the  
supersymmetry allows us to derive more results from the Bethe ansatz  
for finite $N$.  Precisely, we define $w_k$ in terms of $\mu_k$ as
$w_k = (\mu_k-q)/(q\mu_k-1),$  
where $q\equiv\exp(-i\pi/3)$. Then Baxter's Q-function 
\cite{Baxter} ${\cal Q}(w)\equiv\prod_{i=1}^f (w-w_k)$ has
zeroes at $w=w_k$. 
Defining ${\cal R}(w)\equiv {\cal Q}(w)(1+w)^{N-f}$, we find that   
for the $w_k$ giving the ground state,
\begin{equation}  
{\cal R}(q^{-2}w) = t q^{-N}{\cal R}(w) +   
t^{-1} q^N {\cal R}(q^2 w).  
\label{QQQ}  
\end{equation}
We derive an explicit expression for
${\cal R}(w)$ in the sequel \cite{FSii}, but from (\ref{QQQ}) directly
we can rederive $f$ and $t$ for the ground state(s).
When $N=3p$ with $p$ an integer,  
there are non-trivial solutions to (\ref{QQQ}) only when  
$f=N/3$ and $t=(-1)^N\exp(\pm i\pi/3)$. For $N\ne 3p$,   
one has only a single solution with $f=\hbox{int}((N+1)/3)$, and  
$t=(-1)^{N-1}$. 

We can see heuristically why the ground state has $f=\hbox{int}((N+1)/3)$.  
The potential term in (\ref{ham1}) alone is minimized by the state with  
a fermion on every third site; adding any more fermions forces  
fermions to be two sites away and raises the energy. The hopping term  
alone also discourages fermions from being only two sites away,
because it has negative eigenvalues when fermions can hop to an  
adjacent site and back again, and the hard cores prevent this if there  
is another fermion two sites away. The state with a fermion on every  
third site, 
$$\dots 
\bullet\circ\circ
\bullet\circ\circ
\bullet\circ\circ
\bullet\circ\circ
\bullet\circ\circ
\bullet\circ\circ
\bullet\circ\circ
\bullet\circ\circ
\bullet\circ\circ
\dots$$ resembles a N\'eel state for a Heisenberg anti-ferromagnet. 
It is not an eigenstate of the Hamiltonian: the
full ground state is disordered. However, like the N\'eel state, we
expect this state to be a part of the ground state. Our derivations
of $f_{GS}=\hbox{int}((N+1)/3)$ confirm this intuition.
This heuristic picture also gives the fermion numbers of the low-lying
excited states.  The excitations include defects (domain walls) in the
N\'eel-like state, such as
$$\dots 
\bullet\circ\circ
\bullet\circ
\bullet\circ\circ
\bullet\circ\circ
\bullet\circ
\bullet\circ\circ
\bullet\circ\circ
\bullet\circ
\bullet\circ\circ 
\bullet\circ\circ 
\dots$$ The fermion number of this
configuration is just one higher than that of the N\'eel-like state,
and it has three identical defects. Since defects can be moved
arbitrarily far apart with no change in the potential, it is
natural to treat each defect as a quasiparticle with charge $1/3$.
The existence of fractional charge in $1+1$ dimensions is an old
story; this was first discovered in field theory \cite{JRGW}.

Finally, we give the field theory describing the continuum limit of
(\ref{ham1}).  When taking $N$ large, one can rewrite the Bethe equations
in terms of densities of roots \cite{Bethe}, and 
then derive integral equations (known as thermodynamic Bethe
ansatz equations) yielding the free energy. Our model has the same
thermodynamic equations as the XXZ chain at
$\Delta=1/2$, so the two models coincide in the continuum limit.  The
continuum limit of the XXZ chain is described by the massless
Thirring model \cite{hank}, or equivalently a free massless boson
$\Phi$ with action \cite{Friedan}
$$S=\frac{2g}{\pi}\int dx\,dt\ \left[(\partial_t \Phi)^2 -
(\partial_x\Phi)^2\right].$$ The continuum limit of the $\Delta=1/2$
model has $g=2/3$; this is the simplest field theory with ${\cal
N}=(2,2)$ superconformal symmetry \cite{Friedan}.  The $(2,2)$ means
that there are two left and two right-moving supersymmetries: in the
continuum limit the fermion decomposes into left- and right-moving
components over the Fermi sea. The boson also can be decoupled into
left and right pieces, so that $\Phi=\Phi_L + \Phi_R$, while its dual
$\widetilde\Phi=g(\Phi_L-\Phi_R)$.  The states of the field theory are
given by the vertex operators $V_{m,n}= \exp(im \Phi+i
n\widetilde{\Phi})$, of conformal dimensions $h_{L,R}=(m\pm g
n)^2/(4g)$. The four components of the Dirac fermion in the Thirring
model are $V_{\pm 1,\pm 1/2}$, while the supersymmetry generators are
$Q_L^\pm=V_{\pm 1, \pm 3/2}$ and $Q_R^\pm=V_{\pm 1, \mp 3/2}$.  In a
finite size ${\cal L}$, the lowest-energy state is in the
Neveu-Schwarz sector, where the Thirring fermion has anti-periodic
boundary conditions.  This state has $E_{NS}=-\pi/(6{\cal L})$. The
lowest-energy states in the Ramond (periodic boundary conditions)
sector are given by $|\pm\rangle_{R} = V_{0,\pm 1/2}|0\rangle_{NS}$;
both have energy zero. States in this conformal field theory can
be built up by operating with the ``spinons'' $V_{\pm 1/3,\pm 1/2}$.

Comparing this superconformal field theory with the lattice model on
$N=3p$ sites, we identify our two $E=0$ ground states with the two
Ramond vacua; all other states of the lattice model are in the Ramond
sector as well. The $U(1)$ quantum number $m$ corresponds to $f-N/3$
(the fermion number relative to the ground state).  The spinons here
have charge $\pm 1/3$, so it is natural to identify these with the
fractionally-charged excitations in the lattice model.

One can define a quantum dimer model by placing the variables
$c_i^\dagger$ on the links instead of the sites of a lattice.
The product in ${\cal P}_{<i>}$ is then over links which
meet the link $i$, and the projections are precisely
such that overlapping dimers are avoided. On the $N=3p$-site chain, the
ground states have dimer number $N_d=N/3$, which is 2/3 of the
value for close-packed dimers. A more involved example is that of the
supersymmetric dimer model on a 2-leg ladder with $N+1$ rungs. Using
cohomology techniques, we found that the number $N_{GS}$ of
ground states grows quickly with $N$, and that not all
ground states have the same dimer number $N_d$, leading to (partial)
cancellations in the Witten index $W$.  For example, for $N=7$ one
finds five $E=0$ ground states with $N_d=5,5,5,5,6$, so $ W =-3$. Our results, 
asymptotically precise for $N$ large, are $N_{GS} \sim (1.395)^N$ and 
$ W \sim (1.356)^N$. These huge degeneracies suggest additional 
symmetries in the model. It will be interesting to see if supersymmetry 
can be of use in other quantum dimer models \cite{MSF}.
  
Following the ideas above, one can obtain other supersymmetric models
by including more projectors in (\ref{susydef}), or by including
several species of fermions. Consider a
two-species model, with fermions labeled by $+$ and $-$, and with the
conditions that (i) a single site may not be occupied by two particles
and (ii) same-type particles may not occupy nearest neighbor sites. On
a periodic chain with $4n$ sites, the Witten index of this model
turns out to be $W=3$, and we have strong indications that the
continuum limit of this theory is the second model (at $c=3/2$) of the
series of ${\cal N}$=2 minimal superconformal field theories. A
typical ground state pattern is
$$ \ldots 
\circ + \circ - \circ + \circ - \circ + \circ - \circ 
\ldots $$  
and one recognizes the possibility of domain walls of charge $\pm 1/2$ 
($\circ+-\ \circ$) and neutral defects ($+\circ+$). The $+/-$ pattern 
indicates an Ising substructure in the model, in accord with the fact 
that the $c=3/2$ ${\cal N}=2$ minimal model can be written in terms of 
a Majorana fermion and a free boson.   

We finally remark that in our models, the space
of states is made up solely of fermions on which the supersymmetry acts
non-linearly. (This does not preclude a linear realization 
on bosons and fermions in the continuum theory.)
Such realizations of ${\cal N}=2$ lattice
supersymmetry seem very different from other 
known realizations, for example in lattice gauge theory \cite{feo}.

\medskip {\bf Acknowledgements:}
It is a pleasure to thank Bernard Nienhuis, Erik Verlinde and 
Alexander Schr\-ijver for discussions. 
This work was supported in part by the foundations FOM and NWO 
of the Netherlands, by NSF grant DMR-0104799, and by a DOE OJI award.

\end{document}